# MORPHOLOGY AND MAGNETIC PROPERTIES OF $Fe_3O_4$-ALGINIC ACID NANOCOMPOSITES

Małgorzata Kaźmierczak[a,b], Katarzyna Pogorzelec-Glaser[a], Andrzej Hilczer[a], Stefan Jurga[b], Łukasz Majchrzycki[c], Marek Nowicki[c], Ryszard Czajka[c], Filip Matelski[c], Radosław Pankiewicz[d], Bogusława Łęska[d], Leszek Kępiński[e] and Bartłomiej Andrzejewski[a] *

[a]*Institute of Molecular Physics Polish Academy of Sciences, M. Smoluchowskiego 17, PL-60179 Poznan, Poland*
[b]*NanoBioMedical Centre, Adam Mickiewicz University, Umultowska 85, PL-61614 Poznan, Poland*
[c]*Poznań University of Technology, Nieszawska 13A, PL-60965 Poznan, Poland*
[d]*Faculty of Chemistry, Adam Mickiewicz University, Umultowska 89b, PL-61614 Poznan, Poland*
[e]*Institute of Low Temperature and Structure Research, Polish Academy of Sciences, Okolna 2, PL-50422 Wroclaw, Poland*

**Abstract**

Morphology, structure and magnetic properties of nanocomposites of magnetite ($Fe_3O_4$) nanoparticles and alginic acid (AA) are studied. Magnetite $Fe_3O_4$ nanoparticles and the nanoparticles capped with alginic acid exhibit very distinct properties. The chemical bonding between alginic acid and surface of magnetite nanoparticles results in recovery of surface magnetization. On the other hand, it also leads to enhanced surface spin disorder and unconventional behavior of magnetization observed in $Fe_3O_4$-AA nanocomposites at low temperatures.

<u>Keywords</u>: nanocomposite, magnetite nanoparticles, alginic acid, enhanced magnetization

## 1. Introduction

The interest in composites of polymers with magnetic nanoparticles stems from their unique physical properties and potential future applications for magnetic data storage [1], electronic devices and sensors [2], for biomedical applications in magnetic resonance imaging [3], drug delivery [4] and hyperthermia agents [5]. From this point of view, one of the most preferred magnetic materials is magnetite $Fe_3O_4$ because it is a biocompatible mineral with low toxicity (for example crystals of magnetite are magnetoreceptors in the brains of some animal [6]). It also exhibits large magnetic moment and spin polarized electric current - the features highly desired for applications in spintronics.

In bulk, magnetite crystallizes in the inverse spinel $AB_2O_4$ structure with two nonequivalent Fe sites placed in the *fcc* lattice of $O^{2-}$ ions. Tetrahedral A sites contain $Fe^{2+}$ ions, whereas octahedral B site are occupied by $Fe^{2+}$ and $Fe^{3+}$ ions. Magnetic sublattices located on A and B sites are ferrimagneticaly coupled. Mixed valence of Fe ions and fast electron hopping between B sites are responsible for relatively high electric conductivity of $Fe_3O_4$ above the Verwey transition $T_v \approx 125K$ [7].

Nanostructured magnetite exhibits different magnetic, electronic and optical properties than the bulk material. Particularly, a significant reduction in magnetization at the surface of $Fe_3O_4$ nanoparticles makes them useless for many of applications. This obstacle can be overcome by capping the magnetic nanoparticles with polymers [8] or organic acids, which allows to restoration of the surface magnetism [9].

One of the best capping material is alginic acid which is a cheap, common and nontoxic natural biopolymer [10, 11]. The aim of this work is to study the effect of alginic acid

---

*corresponding author: and@ifmpan.poznan.pl

capping on the surface magnetization recovery in Fe$_3$O$_4$ nanoparticles.

## 2. Experimental

### 2.1. Sample Synthesis

All chemicals used in the experiments were purchased from SIGMA ALDRICH. To obtain disaggregated nanoparticles of magnetite Fe$_3$O$_4$, a portion of 9.0 mmol of FeCl$_3$·6H$_2$O was dissolved in 200 mL of ethylene glycol. The solution was vigorously stirred. After 15 min. 131.7 mmol of CH$_3$COONa and 1.575 mmol of polyethylene glycol PEG 400 were added and stirring was continued until they completely dissolved. Then, the solution was transferred into 50 mL teflon reactors and heated using microwave radiation (MARS 5, CEM Corporation) at 160°C for 25 min. The black suspension of nanoparticles obtained as a result of the reaction, was first cooled, isolated by centrifugation and washed with absolute ethanol. The final product was dried in vacuum oven at 40°C. The nanocomposite was prepared from aqueous dispersion of magnetite nanopowder and alginic acid (AA), which was next air dried at room temperature. The nanocomposite had the form of flakes with flat surface.

### 2.2. Sample Characterization

The crystallographic structure of the samples was studied by means of X-ray powder diffraction (XRD) using an ISO DEBYEYE FLEX 3000 instrument with a Co lamp (λ=0.17928 nm). The morphology of Fe$_3$O$_4$ nanoparticles was observed using a Philips CM20 SuperTwin Transmission Electron Microscope (TEM). The structure of nanocomposites was studied by means of Atomic Force Microscope (Dimension Icon®, Bruker) using the Magnetic Force Microscope (MFM) mode and NANO-SENSORS™ PPP-MFMR probes. The magnetic measurements were performed using the Quantum Design Physical Property Measurement System (PPMS) fitted with a Vibrating Sample Magnetometer (VSM) probe.

## 3. Results and discussion

Fig. 1. shows the X-ray powder diffraction patterns of as-obtained Fe$_3$O$_4$ nanoparticles (panel a) and of Fe$_3$O$_4$-AA nanocomposite with magnetite content equal to 10% in weight (panel b).

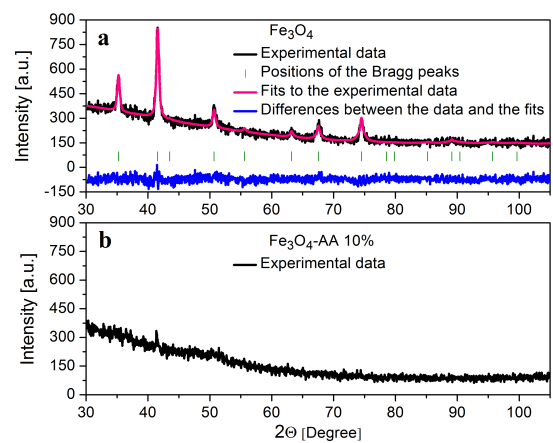

Fig. 1. XRD powder pattern and line profile fitting of Fe$_3$O$_4$ nanoparticles a) and Fe$_3$O$_4$-AA nanocomposite b).

The solid line corresponds to the best Rietveld profile fit calculated by means of FULLPROF software for the cubic crystal structure with $Fd\bar{3}m$ space group and X-ray radiation of the wavelength 0.17928 nm as used in the experiment. The vertical bars correspond to the Bragg peaks and the line below them is the difference between the experimental data and the fit. XRD studies verified the $Fd\bar{3}m$ point group of the Fe$_3$O$_4$ nanopowder with lattice parameters $a$=8.3641 Å and mean crystallite size 20 nm determined using the Scherrer method. For the composite, the intensity of diffraction peaks is too low to perform any analysis, even if the content of magnetite is high and equal to 10% in weight.

The TEM image of magnetite nanoparticles is presented in the inset to Fig. 2. The magnetic nanoparticles are almost monodisperse and spherical. The in size distribution of Fe$_3$O$_4$ nanoparticles (histogram) is presented in Fig. 2. The distribution can be fitted by the log-normal function:

$$f(x) = \frac{1}{x\sqrt{2\pi\sigma^2}} \exp\left[-\frac{1}{2\sigma^2}\ln^2\left(\frac{x}{\langle x \rangle}\right)\right] \qquad (1)$$

where: ⟨x⟩ is the mean size of nanoparticles and σ is distribution width. The values characterizing the distribution are: ⟨x⟩=20.5nm and σ=0.11, with ⟨x⟩ corresponding well to XRD data.

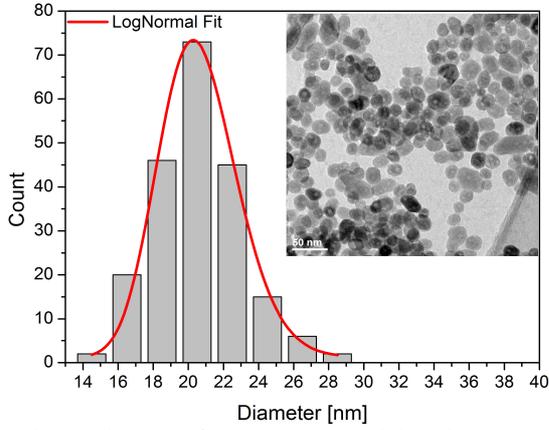

Fig. 2. Histogram for $Fe_3O_4$ nanoparticles with a log-normal fitting. The TEM image is shown in the inset.

Fig. 3 shows the topography (panel a), elastic properties (panel b) and magnetic domains (panels c and d) of the $Fe_3O_4$-AA composite surface with $Fe_3O_4$ content 10% in weight studied by MFM.

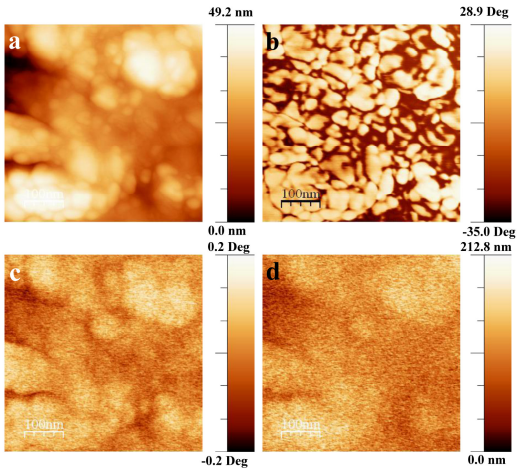

Fig. 3. a) surface topography, b) phase contrast, c) magnetic phase and d) magnetic amplitude images for $Fe_3O_4$-AA composite with $Fe_3O_4$ content 10% in weight.

The roughness of the surface is below 10 nm for the scanned area of 500×500 nm. The knobs on the topography image (white spots) indicate the presence of small agglomerates of $Fe_3O_4$ nanoparticles, which are also seen as a white areas at the phase contrast image (panel b). The amplitude and phase contrast of the magnetic signal (panels c and d) indicates the presence of magnetic domains of the size close to 100 nm. The actual size of these domains can be smaller than that presented in the figures because of insufficient spatial resolution of the MFM method (about 50 nm) which causes smearing of the images.

The results of magnetic study are presented in Figs. 4 and 5. The magnetization is normalized with respect to the content of magnetite in the samples. For $Fe_3O_4$ nanoparticles, the temperature dependence of magnetization $M \sim T^{1.9}$ deviates from the Bloch law $M \sim T^{1.5}$ which is valid for the capped nanoparticles of magnetite (see Fig. 4). The deviation from the Bloch law for uncapped $Fe_3O_4$ nanoparticles can be related to degraded magnetic ordering at the surface. Moreover, the magnetization of $Fe_3O_4$ nanoparticles at room temperature is only 51 $Am^2$/kg, i.e. much below the saturation value for bulk magnetite (~90 $Am^2$/kg) and also lower than the magnetization of the capped particles, equal to 60 $Am^2$/kg. The enhancement of magnetization and the Bloch-like behavior of the capped nanoparticles can be explained in terms of recovery of surface magnetism due to the chemical bonding between AA and $Fe_3O_4$ nanoparticles. This bonding between O atoms in the carboxylic groups and two of four Fe atoms in the surface unit cell Fe-O, makes the coordinations and distances close to those in bulk [9]. The rest two Fe atoms still exhibit reduced magnetization because they are closer to the in-plane oxygens, which results in partially empty $d_{x^2-y^2}$ orbitals. The inhomogeneity in the Fe coordination can be responsible for increased spin disorder or unconventional magnetism at the $Fe_3O_4$ surface. This unconventional behavior is manifested as a rapid increase in magnetization at low temperature observed for $Fe_3O_4$-AA composites (see Fig 4). The alternative explanation of this magnetization upturn assumes quantization of spin-wave spectrum due to the finite size of the particles which occurs at low temperatures and is responsible for the deviation from the Bloch law [12].

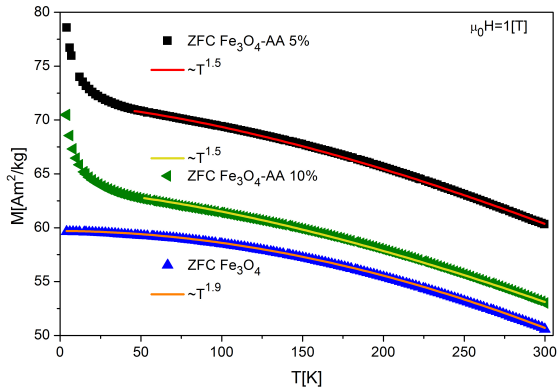

Fig. 4. Magnetization $M(T)$ for $Fe_3O_4$ nanoparticles and $Fe_3O_4$-AA composites containing 5% and 10% of magnetite in weight.

The magnetization loops $M(H)$ for $Fe_3O_4$ nanoparticles and $Fe_3O_4$–AA composites are shown in Fig. 5.

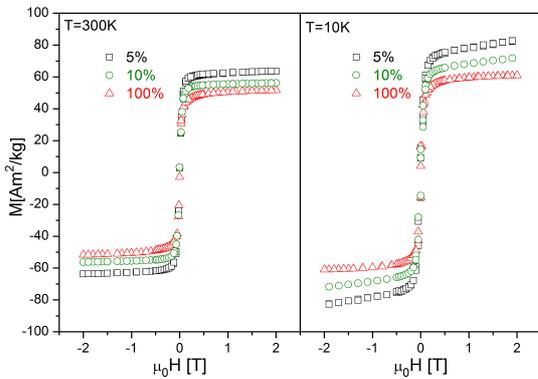

Fig. 5. Magnetization loops $M(H)$ for $Fe_3O_4$ nanoparticles and $Fe_3O_4$-AA composites with 5% and 10% of magnetite content.

Both the nanoparticles and composites exhibit ferromagnetic (ferrimagnetic) hysteresis loops, which saturate above about 0.3 T. The magnetization of the composites is enhanced as compared to that of the uncapped $Fe_3O_4$ nanoparticles. At low temperatures the magnetization loops for the composites are superposition of ferromagnetic and linear contribution from unconventional magnetism. This unconventional behavior cannot be simply related to the paramagnetism at the degraded $Fe_3O_4$ surface because it is absent in the uncapped nanoparticles of magnetite.

## Conclusions

Capping of $Fe_3O_4$ nanoparticles with alginic acid leads to partial recovery of surface magnetization. On the other hand, bonding between alginic acid and $Fe_3O_4$ nanoparticles by means of O atoms results in unconventional magnetism observed at low temperatures.

## Acknowledgments

This project has been supported by National Science Centre by the project No. N N507 229040. M.K. was supported through the European Union - European Social Fund and Human Capital - National Cohesion Strategy.